\documentclass[aps,twocolumn,groupedaddress,amsmath,amssymb]{revtex4}
\usepackage{amsmath, dsfont}
\usepackage{amssymb}
\usepackage[dvips]{graphics}
\usepackage{epsfig}
\usepackage{calc}
\usepackage{amsfonts}
\usepackage{graphicx}
\usepackage[ansinew]{inputenc}

\begin{document}
\setcounter{page}{1}
\title[]{Thermodynamics of a BTZ black hole solution with an Horndeski source}

\author{Moises Bravo-Gaete}\email{mbravog-at-inst-mat.utalca.cl}
\affiliation{Instituto de Matem\'atica y F\'{\i}sica, Universidad de
Talca, Casilla 747, Talca, Chile.}

\author{Mokhtar Hassaine}\email{hassaine-at-inst-mat.utalca.cl}
\affiliation{Instituto de Matem\'atica y F\'{\i}sica, Universidad de
Talca, Casilla 747, Talca, Chile.}

\begin{abstract}
In three dimensions, we consider a particular truncation of the
Horndeski action that reduces to the Einstein-Hilbert Lagrangian
with a cosmological constant $\Lambda$ and a scalar field whose
dynamics is governed by its usual kinetic term together with a
nonminimal kinetic coupling. Requiring the radial component of the
conserved current to vanish, the solution turns out to be the BTZ black hole geometry with
a radial scalar field well-defined at the horizon. This means in particular that the stress tensor associated to the matter source
behaves on-shell as an effective cosmological constant term. We
construct an  Euclidean action whose field equations are consistent
with the original ones and such that the constraint on the radial
component of the conserved current also appears as a field equation. With
the help of this Euclidean action, we derive the mass and the
entropy of the solution, and found that they are proportional to the
thermodynamical quantities of the BTZ solution by an overall factor that depends
on the cosmological constant. The reality condition and the
positivity of the  mass impose  the cosmological constant to be
bounded from above as $\Lambda\leq-\frac{1}{l^2}$ where the limiting case
$\Lambda=-\frac{1}{l^2}$  reduces to the BTZ solution with a vanishing scalar
field. Exploiting  a scaling symmetry of the reduced action, we also
obtain the usual three-dimensional Smarr formula.  In the last section, we extend all these results in higher dimensions
where the metric turns out to be the Schwarzschild-AdS spacetime with planar horizon.

\end{abstract}

\maketitle

\section{Introduction}
Since the  discovery of the BTZ black hole solution
\cite{Banados:1992wn}, three-dimensional Einstein gravity with a
negative cosmological constant has become  an important field of
investigations. A considerable number of papers has been devoted to
the physical and mathematical implications of the BTZ solution in
particular in  the context of AdS$_{3}$/CFT$_{2}$ correspondence.
Indeed, it is now well accepted that three-dimensional gravity is an
excellent laboratory in order to explore and test some of the ideas
behind the AdS/CFT correspondence \cite{Maldacena:1997re}.

It is well-known that the static BTZ geometry  whose line element
is given by
\begin{eqnarray}
ds^2=-\Big(\frac{r^2}{l^2}-M\Big)dt^2+\frac{dr^2}{\frac{r^2}{l^2}-M}+r^2d\varphi^2,
\label{BTZmetric}
\end{eqnarray}
is a solution of the Einstein equations with a fixed value of the
negative cosmological constant $-{l^{-2}}$,
$$
G_{\mu\nu}-{l^{-2}}g_{\mu\nu}=0.
$$
Here, we will exhibit a matter action that sources the BTZ
spacetime. In order to achieve this task, the corresponding stress
tensor $T_{\mu\nu}$ of the matter source must behave on-shell as an
effective cosmological constant term, i. e.
$$
T^{\mbox{\tiny{on-shell}}}_{\mu\nu}=\left({l^{-2}}+\Lambda\right)g_{\mu\nu}.
$$
Indeed, in this case, it is simple to realize that the Einstein
equations
$$
G_{\mu\nu}+\Lambda g_{\mu\nu}=T_{\mu\nu},
$$
will automatically be satisfied on the BTZ metric (\ref{BTZmetric}).
For that purpose, we consider the following three-dimensional action
{\small\begin{eqnarray}
 S=\int\sqrt{-g}\,d^{3}x\left(R-2\Lambda
  -\frac{1}{2}\left(  \alpha g_{\mu\nu}-\eta
G_{\mu\nu}\right)
\nabla^{\mu}\phi\nabla^{\nu}\phi\right),\label{action1}
\end{eqnarray}}
where $R$ and $G_{\mu\nu}$ stand respectively for the Ricci scalar
and the Einstein tensor. This model is part of the so-called
Horndeski action which is the most general tensor-scalar action
yielding at most to second-order field  equations in four dimensions
\cite{Horndeski:1974wa}. In three dimensions, the corresponding
field equations are also of second order, and are given by
\begin{subequations}
\label{eqs}
\begin{eqnarray}
\label{eqmetric}
&&G_{\mu\nu}+\Lambda g_{\mu\nu}=\frac{1}2\left[{\alpha} {T}_{\mu \nu}^{(1)}+{\eta} {T}_{\mu \nu}^{(2)}\right],\\
\nonumber\\
\label{eqphi} &&\nabla_{\mu}\left[  \left(  \alpha g^{\mu\nu}-\eta
G^{\mu\nu}\right) \nabla_{\nu}\phi\right]  =0,
\end{eqnarray}
\end{subequations}
where the stress tensors $T_{\mu\nu}^{(i)}$ are defined by {\small
\begin{eqnarray} \label{stresstensor} &&T_{\mu\nu}^{(1)}
=\Big(\nabla_{\mu}\phi\nabla_{\nu}\phi-\frac
{1}{2}g_{\mu\nu}\nabla_{\lambda}\phi\nabla^{\lambda}\phi \Big).\\
&&T_{\mu\nu}^{(2)}= \frac{1}{2}\nabla_{\mu}\phi\nabla_{\nu }\phi
R-2\nabla_{\lambda}\phi\nabla_{(\mu}\phi R_{\nu)}^{\lambda}
-\nabla^{\lambda}\phi\nabla^{\rho}\phi R_{\mu\lambda\nu\rho}\nonumber\\
&&-(\nabla_{\mu}\nabla^{\lambda}\phi)(\nabla_{\nu}\nabla_{\lambda}%
\phi)+(\nabla_{\mu}\nabla_{\nu}\phi)\square\phi+\frac{1}{2}G_{\mu\nu}%
(\nabla\phi)^{2}\nonumber\\
&&-g_{\mu\nu}\left[ -\frac{1}{2}(\nabla^{\lambda}\nabla^{\rho}\phi
)(\nabla_{\lambda}\nabla_{\rho}\phi)+\frac{1}{2}(\square\phi)^{2}%
-\nabla_{\lambda}\phi\nabla_{\rho}\phi
R^{\lambda\rho}\right].\nonumber
\end{eqnarray}}
The scalar field equation (\ref{eqphi}) is a current conservation
equation which is a consequence of the shift symmetry of the action,
$\phi\to\phi+\mbox{const.}$.

The first exact black hole solution of
these equations without cosmological constant was found in
\cite{Rinaldi:2012vy}. However, in this case, the scalar field
becomes imaginary outside the horizon. Recently, this problem has
been circumvented by adding a cosmological constant term yielding to
asymptotically locally (A)dS (and even flat for $\alpha=0$)  black
hole solutions with a real scalar field outside the horizon
\cite{Anabalon:2013oea}. The electric charged version of the AdS
solutions with a Maxwell field have been studied in
\cite{Cisterna:2014nua}. The field equations (\ref{eqs}) admit other
interesting solutions with a nontrivial and regular time-dependent
scalar field on a static and spherically symmetric spacetime
\cite{Babichev:2013cya}. Interestingly enough, this solution in the
particular case of $\Lambda=\eta=0$ reduces to an unexpected stealth
configuration on the Schwarzschild metric \cite{Babichev:2013cya}.
There also exist Lifshitz black hole
solutions with a time-independent scalar field for a fixed value of  the dynamical exponent
$z=\frac{1}{3}$, \cite{Bravo-Gaete:2013dca}. Solutions for a more general
truncation of the Horndeski action that is shift-invariant as well
as enjoying the reflection symmetry $\phi\to -\phi$ have been
obtained in \cite{Kobayashi:2014eva}. In all these examples, in
order to satisfy the radial part of the current conservation
(\ref{eqphi}) without imposing any constraints on the radial
derivative of the scalar field, the geometry has been chosen such that
\begin{eqnarray}
\alpha g^{rr}-\eta G^{rr}=0. \label{cond}
\end{eqnarray}
We note that this condition simplifies considerably the field
equations in particular in the time-independent case where the full
conservation equation (\ref{eqphi}) is automatically satisfied
without constraining the radial dependence of the scalar field.

In the present work, we will show that the BTZ geometry naturally
emerges as a solution of this particular Horndeski action
(\ref{action1}) in  three dimensions subjected to the condition
(\ref{cond}). We will also
 analyze the thermodynamical implications of such solution and extend all our results in arbitrary dimension. The plan of the paper is organized as follows. In the next section,
we present in details the derivation of the solution using the constraint
(\ref{cond}). In Sec. III,  we construct an Euclidean
action whose field equations turn to be consistent with the original
ones and such that the constraint on the radial current (\ref{cond})
naturally appears as a field equation. This construction will be
useful to obtain the mass and the entropy of the solution. The usual
Smarr formula is also derived by exploiting a scaling symmetry of
the reduced action. The rotating version of the solution as well as
a particular example of time-dependent solution will be reported in
Sec. IV. In Sec. V, we extend all the results to arbitrary dimension where the metric solution is nothing but the Schwarzschild-AdS spacetime with planar horizon.
The last section is devoted to our conclusions.

\section{Derivation of the solution}
Let us derive the most general   solution of the field equations
(\ref{eqs}) subjected to the condition (\ref{cond}) with an Ansatz of the form
\begin{eqnarray}
\label{ansatz}
&&ds^2=-h(r)dt^2+\frac{dr^2}{f(r)}+r^2d\varphi^2,\nonumber\\
&&\phi=\phi(r).
\end{eqnarray}
For clarity, we define
$$
\epsilon_{\mu\nu}:=G_{\mu\nu}+\Lambda
g_{\mu\nu}-\frac{1}2\left[{\alpha} {T}_{\mu \nu}^{(1)}+{\eta}
{T}_{\mu \nu}^{(2)}\right].
$$
The condition (\ref{cond}) on the geometry becomes
\begin{eqnarray}
f(r)=\frac{2\alpha r h(r)}{\eta h^{\prime}(r)},
\end{eqnarray}
and automatically implies that the current conservation
(\ref{eqphi}) is satisfied. Using this last relation, the radial
component of the Einstein equations $\epsilon_{rr}=0$ allows to
express the square of the derivative of the scalar field as
$$
\left(\phi^{\prime}\right)^2=-\frac{(\alpha+\eta\Lambda)h^{\prime}}{\alpha^2
r h}.
$$
The remaining independent Einstein equation, $\epsilon_{tt}=0$ or
equivalently $\epsilon_{\varphi\varphi}=0$, yields
$$
\epsilon_{tt}\propto
\left(\alpha-\eta\Lambda\right)\left[rh^{\prime\prime}-h^{\prime}\right]=0.
$$
Hence, it is clear that the point defined by $\alpha=\eta\Lambda$
corresponds to a degenerate sector \cite{Bravo-Gaete:2013dca}, while
for  $\alpha\not=\eta\Lambda$, the solution is given by
$$
h(r)=Cr^2-M,\qquad f(r)=\frac{\alpha}{\eta C}(Cr^2-M),
$$
where $C$ and $M$ are two integration constants. In order to deal with
the BTZ metric (\ref{BTZmetric}), we choose $C=l^{-2}$ and the
coupling constants must be fixed such that
\begin{eqnarray}
\label{alphaeta}
\frac{\alpha}{\eta}=l^{-2}.
\end{eqnarray}
Note that the degenerate sector $\alpha=\eta\Lambda$ corresponds to
a choose of the cosmological constant
\begin{eqnarray}
\Lambda^{\mbox{\tiny{degenerate}}}=l^{-2}.
\end{eqnarray}
In sum, for $\frac{\alpha}{\eta}=l^{-2}$ and
$\Lambda\not=\Lambda^{\mbox{\tiny{deg}}}$, the solution is given by
the BTZ metric (\ref{BTZmetric}) together with a radial scalar field
\begin{subequations}
\label{sol}
\begin{eqnarray}
&&\xi(r):=\left(\phi^{\prime}\right)^2=-\frac{2(\Lambda l^2+1)}{\eta\left(\frac{r^2}{l^2}-M\right)}\\
&&\phi(r)=\pm \sqrt{-\frac{2\,l^{2}(\Lambda
l^2+1)}{\eta}}\ln\left(\frac{r}{l}+\sqrt{\frac{r^2}{l^2}-M}\right),
\end{eqnarray}
\end{subequations}
provided that
\begin{eqnarray}
\label{rc} \frac{2\,l^{2}(\Lambda l^2+1)}{\eta}\leq 0.
\end{eqnarray}
Various comments can be made concerning this solution. Firstly, we
note that for $\Lambda=-l^{-2}$, the scalar field vanishes
identically and the solution reduces to the BTZ solution. Secondly, the scalar field is
well-defined at the horizon $r_+=l\sqrt{M}$, and as expected the
stress tensor of the matter part behaves on-shell as an effective
cosmological constant
$$
\frac{1}2\left[{\alpha} {T}_{\mu \nu}^{(1)}+{\eta} {T}_{\mu
\nu}^{(2)}\right]^{\mbox{\tiny{on-shell}}}=\left(\Lambda+l^{-2}\right)g_{\mu\nu}.
$$
As a last comment, we remark that for $\alpha>0$ which corresponds to the right sign of the standard kinetic term, the previous reality conditions (\ref{alphaeta}-\ref{rc}) will
imply that $\eta>0$ and the cosmological constant $\Lambda$ must be bounded from above
as $\Lambda\leq -l^{-2}$. The limiting case $\Lambda= -l^{-2}$ corresponding to the BTZ solution without source.

In what follows, we will derive the mass and the entropy of the solution (\ref{sol}).

\section{Thermodynamics of the black hole solution}
The partition function for a thermodynamical ensemble is identified
with the Euclidean path integral in the saddle point approximation
around the Euclidean continuation of the classical solution
\cite{Gibbons:1976ue}. The Euclidean and Lorentzian action are
related by $I_{E}=-iI$ where the periodic Euclidean time is  $\tau
=it$. The Euclidean continuation of the class of metrics considered
here  is given by \footnote{ For the Ansatz considered in
(\ref{ansatz}), this will correspond to $h(r)=N(r)^2F(r)$ and
$f(r)=F(r)$.}
$$
ds^2=N^2(r)\,F(r)d\tau^2+\frac{dr^2}{F(r)}+r^2d\varphi^2.
$$
In order to avoid conical singularity at the horizon in the
Euclidean metric, the Euclidean time is made periodic with period
$\beta$ and the Hawking temperature $T$ is given by $T=\beta^{-1}$.
Since we are only interested  in  static solution with a radial
scalar field, it is enough to consider a \textit{reduced}
action principle. However, there is an important subtlety that has to do with the constraint (\ref{cond}) we used in order to derive our solution. Indeed, this constraint
together with the fact of looking for a static scalar field make the equation associated to the variation of the scalar field (\ref{eqphi}) redundant
in the sense that the equation is automatically satisfied.
 Hence, in our reduced action, the constraint (\ref{cond}) should appear as a field equation in order to deal with an equivalent problem. This can be achieved considering
 the following Euclidean action
\begin{widetext}
\begin{eqnarray}
\label{redaction} I_E:=I_E(N,F,\xi)=&&2\pi \beta\int_{r_+}^{\infty}
N\left[F^{\prime}+2\Lambda r+\frac{\alpha}{2}rF\xi+\frac{3}{4}\eta
FF^{\prime}\xi+\frac{\eta}{2}F^2\xi^{\prime}\right]dr+B_E,
\end{eqnarray}
\end{widetext}
where the dynamical field is chosen to be $\xi(r):=(\phi^{\prime})^2$  and not the scalar field itself $\phi$. Note that $r_+$ is the location of the horizon and $B_E$ is a boundary
term that is fixed by requiring that the Euclidean action has an
extremum, that is $\delta I_{E}=0$. In this case, the variation with respect to the dynamical fields $N, F$ and $\xi$ yield
\begin{eqnarray}
\label{eqsE}
&&E_N:=F^{\prime}+2\Lambda r+\frac{\alpha}{2}rF\xi+\frac{3}{4}\eta FF^{\prime}\xi+\frac{\eta}{2}F^2\xi^{\prime}=0,\nonumber\\
&&E_F:=-N^{\prime}\left(1+\frac{3}{4}\eta F\xi\right)+N\left(\frac{\alpha}{2}r\xi+\frac{1}{4}{\eta}F\xi^{\prime}\right)=0,\nonumber\\
&&E_{\xi}:=-\frac{\eta}{2}N^{\prime}F^2+N\left(\frac{\alpha}{2}r
F-\frac{1}{4}\eta F^{\prime}F\right)=0,
\end{eqnarray}
and the last equation $E_{\xi}=0$ is
nothing but the constraint used previously (\ref{cond}) to obtain
our solution. At the special point
$\alpha=\frac{\eta}{l^2}$, the equations (\ref{eqsE}) turn out to be
equivalent to the original ones supplemented by the constraint
(\ref{cond}). Indeed, the most general solution of the system
(\ref{eqsE}) can be derived as follows. For $X(r):=4+3\eta
F(r)\xi(r)\not=0$ \footnote{For $X(r)=0$, one ends with a particular case of (\ref{sol}) with a fixed value of the cosmological constant $\Lambda=-\frac{1}{3\,l^2}$}, we consider the combination
$$
-\frac{2\eta F^2}{X}E_F+\frac{\eta N F}{X}E_N+E_{\xi}=0,
$$
which permits to obtain
$$
\xi(r)=-\frac{2(\Lambda l^2+1)}{\eta F(r)}.
$$
Injecting this expression into $E_N=0$, one obtains that
$F(r)=r^2/l^2-M$ where $M$ is an integration constant, and finally
the equation $E_{\xi}=0$ implies that $N$ is constant which can be chosen to $1$
without any loss of generality. Hence, at $\alpha=\frac{\eta}{l^2}$,
the most general solution of the system (\ref{eqsE}) is given by
\begin{eqnarray}
N(r)=1,\quad F(r)=\frac{r^2}{l^2}-M,\quad \xi(r)=-\frac{2(\Lambda
l^2+1)}{\eta F(r)}, \label{solEuc}
\end{eqnarray}
and corresponds to the solution obtained previously (\ref{sol}). We now determine the boundary term of the Euclidean action which is
given by
\begin{eqnarray}
\delta B_E=-2\pi\beta\left[\delta F\left(1+\frac{3}{4}\eta
F\xi\right)+\frac{\eta}{2}F^2\delta\xi\right]_{r_+}^{\infty}.
\end{eqnarray}
In order to obtain $\delta B_E$, we need the variations of the field
solutions (\ref{solEuc}) at infinity
$$
\delta F|_{\infty}=-\delta M,\qquad
(F^2\delta\xi)|_{\infty}=\frac{2(\Lambda l^2+1)}{\eta}\delta
F|_{\infty},
$$
while at the horizon, they are given by
\begin{eqnarray*}
&&\delta F|_{r_+}=-F^{\prime}|_{r_+}\delta r_+=-\frac{4\pi}{\beta}\delta r_+,\\
&&(F^2\delta\xi)|_{r_+}=\frac{2(\Lambda l^2+1)}{\eta}\delta
F|_{r_+}=-\frac{2(\Lambda l^2+1)}{\eta}\frac{4\pi}{\beta}\delta r_+.
\end{eqnarray*}
Hence, we have
$$
I_E=B_E(\infty)-B_E(r_+)=2\pi\Big[\beta M-4\pi
r_+\Big]\left(\frac{1-\Lambda l^2}{2}\right),
$$
and, we can identify the mass ${\cal M}$ and the entropy ${\cal S}$ of the solution to be given
by
$$
{\cal M}=\frac{\partial I_E}{\partial \beta},\qquad {\cal S}=\beta\frac{\partial I_E}{\partial \beta}-I_E,
$$
yielding
\begin{eqnarray}
{\cal M}=2\pi M\left(\frac{1-\Lambda l^2}{2}\right),\qquad {\cal S}=8\pi^2
\left(\frac{1-\Lambda l^2}{2}\right)r_+.
\end{eqnarray}
Since the Hawking temperature is given by $T=\frac{1}{2\pi}r_+$, it
is easy to see that the first law  $d{\cal M}=Td{\cal S}$ holds.  In the BTZ  case $\Lambda=
-l^{-2}$, the scalar field vanishes and the mass and entropy reduce
to the thermodynamical quantities of the BTZ solution \cite{Banados:1992wn}.

We can now go further by exploiting a scaling symmetry of the
reduced action in order to obtain the usual three-dimensional Smarr
formula in the same lines as those done in Ref. \cite{Banados:2005hm}. In fact, it is
easy to see that the reduced action (\ref{redaction}) enjoys the
following scaling symmetry
\begin{eqnarray}
\bar{r}=\sigma r,\quad \bar{N}(\bar{r})=\sigma^{-2}N(r),\nonumber\\
\bar{F}(\bar{r})=\sigma^{2}F(r),\quad\bar{\xi}(\bar{r})=\sigma^{-2}\xi(r),
\end{eqnarray}
from which one can derive a Noether  quantity
\begin{eqnarray*}
C(r)=N\Big[\left(1+\frac{3}{4}\eta F\xi\right)\left(-2F+rF^{\prime}\right)+\frac{\eta}{2}F^2(r\xi^{\prime}+2\xi)\Big],\nonumber\\
\end{eqnarray*}
which is conserved $C^{\prime}(r)=0$ by virtue of the field equations
(\ref{eqsE}). Evaluating this quantity at infinity and at the
horizon, one gets
$$
C(r=\infty)=M(1-\Lambda l^2),\quad
C(r=r_+)=\frac{4\pi}{\beta}r_+\left(\frac{1-\Lambda l^2}{2}\right).
$$
Since the Noether charge is conserved, these expressions must be equal
$$
M(1-\Lambda l^2)=\frac{4\pi}{\beta}r_+\left(\frac{1-\Lambda
l^2}{2}\right),
$$
which in turn implies the following Smarr formula
\begin{eqnarray}
{\cal M}=\frac{T}{2}{\cal S}. \label{Smarr}
\end{eqnarray}
This latter corresponds to the standard three-dimensional Smarr
formula \cite{Smarr}.

\section{Rotating, time dependent  and stealth solutions}
Operating a Lorentz boost in the plane $(t,\varphi)$, we obtain the
rotating version of the solution found previously. At the point
$\alpha=\eta \, l^{-2}$, the metric function turns out to be the
rotating BTZ
\begin{eqnarray}
ds^2=-F(r)dt^2+\frac{dr^2}{F(r)}+r^2\left(d\varphi-\frac{J}{2r^2}dt\right)^2,
\label{BTZrotating}
\end{eqnarray}
where the structural function $F$ is given by
\begin{eqnarray}
F(r)=\frac{r^2}{l^2}-M+\frac{J^2}{4r^2}, \label{F}
\end{eqnarray}
and, where $J$ corresponds to the angular momentum. The scalar field
solution read
\begin{eqnarray}
\xi(r):=(\phi^{\prime}(r))^2=-\frac{2(l^2\Lambda+1)}{\eta F(r)}.
\end{eqnarray}

We also report two solutions with a linear time scalar field on the rotating BTZ metric (\ref{BTZrotating}). The
first one is obtained for  $\alpha=\eta
\,l^{-2}$ and given by
\begin{eqnarray}
\phi(t,r)=q\,t \pm \int \sqrt{\frac{q^2\eta-2F(r)(\Lambda
l^2+1)}{\eta F(r)^2}}dr,
\end{eqnarray}
where $q$ is a constant. The second solution corresponds to a
stealth configuration, that is a solution where both side of the Einstein
equations (\ref{eqmetric}) vanish identically
\begin{eqnarray}
\label{stealtheq} G_{\mu\nu}+\Lambda
g_{\mu\nu}=0=\frac{1}2\left[{\alpha} {T}_{\mu \nu}^{(1)}+{\eta}
{T}_{\mu \nu}^{(2)}\right].
\end{eqnarray}
In fact, for $\Lambda=-1/l^2$, a solution of the stealth equations
(\ref{stealtheq}) is given by the rotating BTZ metric
(\ref{BTZrotating}) together with a time-dependent scalar field
\begin{eqnarray}
\phi(t,r)=q\left(t \pm \int\frac{dr}{F(r)}\right),
\end{eqnarray}
where $q$ is a constant. This stealth solution is different from the
one derived in \cite{AyonBeato:2004ig}, where in this reference the
authors considered as a source a scalar field nonminimally coupled.
Moreover, in \cite{AyonBeato:2004ig}, the time-dependent stealth only exists in the case of the static BTZ metric, that is for $J=0$.

\section{Extension to higher dimensions}
We now extend our analysis in arbitrary $D$ dimensions with the action
{\small\begin{eqnarray}
 S=\int\sqrt{-g}\,d^{D}x\left(R-2\Lambda
  -\frac{1}{2}\left(  \alpha g_{\mu\nu}-\eta
G_{\mu\nu}\right)
\nabla^{\mu}\phi\nabla^{\nu}\phi\right),\label{action2}
\end{eqnarray}}
for which the field equations are given by (\ref{eqs}). Here, we will consider an Ansatz metric with a planar horizon and a static radial
scalar field
\begin{eqnarray*}
&&ds^2=-N^2Fdt^2+\frac{dr^2}{F}+r^2d\vec{x}_{D-2}^2\\
&&\phi=\phi(r).
\end{eqnarray*}
The solution of the field equations for this Ansatz subjected to the constraint (\ref{cond}) is now given by the Schwarzschild-AdS metric with a planar horizon
\begin{subequations}
\label{sol2}
\begin{eqnarray}
&&ds^2=-F(r)dt^2+\frac{dr^2}{F(r)}+r^2d\vec{x}_{D-2}^2,\\
\label{sol2a}
&&F(r)=\frac{r^2}{l^2}-\frac{M}{r^{D-3}},\\
\label{sol2b}
&&\xi(r):=(\phi^{\prime})^2=-\frac{2(2l^2\Lambda+(D-1)(D-2))}{\eta (D-1)(D-2)F(r)},
\label{sol2c}
\end{eqnarray}
\end{subequations}
provided that
\begin{eqnarray}
\label{alphaeta2}
\frac{\alpha}{\eta}=\frac{(D-1)(D-2)}{2l^2}.
\end{eqnarray}
Note that as in the three-dimensional case, one can obtain an explicit expression of the scalar field
\begin{eqnarray*}
\phi(r)&=&\pm \frac{2}{(D-1)}\,
\sqrt{-\frac{2\,l^{2}\big[2\,\Lambda\,l^{2}+(D-1)(D-2)\big]}{\eta\,(D-1)(D-2)}}\nonumber\\
&\times&\ln\left[r^{\frac{D-3}{2}}\,\left(\frac{r}{l}+\sqrt{\frac{r^2}{l^2}-\frac{M}{r^{D-3}}}\right)\right].
\end{eqnarray*}
In the case $D=3$, the solution reduces to the one previously derived (\ref{sol}) on the BTZ spacetime, and in $D=4$, this solution was already reported in Ref. \cite{Anabalon:2013oea}. As before, the stress tensor associated to the variation of the matter source behaves on-shell as an effective cosmological constant term, that is
$$
\frac{1}2\left[{\alpha} {T}_{\mu \nu}^{(1)}+{\eta} {T}_{\mu
\nu}^{(2)}\right]^{\mbox{\tiny{on-shell}}}=\left(\Lambda+\frac{(D-1)(D-2)}{2l^2}\right)g_{\mu\nu}.
$$
The reality condition (\ref{sol2c}) together with the relation (\ref{alphaeta2}) and requiring the standard kinetic term to have the right sign $\alpha>0$ impose the cosmological constant $\Lambda$ to be bounded from above as
\begin{eqnarray}
\Lambda\leq -\frac{(D-1)(D-2)}{2l^2}.
\end{eqnarray}
The Euclidean action is now given by
\begin{widetext}
\begin{eqnarray}
\label{redaction2} I_E(N,F,\xi)=\beta\,\mbox{Vol}(\Sigma_{D-2})&&\int_{r_+}^{\infty}
N\Big[(D-2)r^{D-3}F^{\prime}+2\Lambda r^{D-2}+\frac{\alpha}{2}r^{D-2}F\xi+\frac{3(D-2)}{4}r^{D-3}\eta
FF^{\prime}\xi\nonumber\\
&&+\frac{(D-2)\eta}{2}r^{D-3}F^2\xi^{\prime}+(D-2)(D-3)\left(r^{D-4}F+\frac{\eta}{4}F^2\xi r^{D-4}\right)\Big]dr+B_E,
\end{eqnarray}
\end{widetext}
where $\mbox{Vol}(\Sigma_{D-2})$  stands for the volume of the compact $(D-2)-$dimensional planar manifold, and $r_+=(l^2M)^{1/(d-1)}$ is the location of the horizon. The variation with respect to the dynamical fields $N, F$ and $\xi$ yield
\begin{eqnarray*}
\label{eqsE2}
&&E_N:=(D-2)r^{D-3}F^{\prime}+2\Lambda r^{D-2}+\frac{\alpha}{2}r^{D-2}F\xi\nonumber\\
&&+\frac{3(D-2)}{4}r^{D-3}\eta
FF^{\prime}\xi+\frac{(D-2)\eta}{2}r^{D-3}F^2\xi^{\prime}\nonumber\\
&&+(D-2)(D-3)\left(r^{D-4}F+\frac{\eta}{4}F^2\xi r^{D-4}\right)=0,
\end{eqnarray*}
\begin{eqnarray*}
&&E_F:=- N' \left( (D-2)\,{r}^{D-3}   + \frac{3}{4} (D-2) \, r^{D-3} \eta F \xi \right) \nonumber\\
&& + N \left( \frac{\alpha}{2} r^{D-2} \xi + \frac{(D-2)}{4} r^{D-3} \eta F \xi' \right.\nonumber \\
&& \left. - \frac{(D-3)(D-2)}{4} r^{D-4} F \eta \xi \right) =0,\nonumber\\
\end{eqnarray*}
\begin{eqnarray*}
&&E_{\xi}:=- \frac{(D-2)}{2} \eta \, r^{D-3} \, N' \, F^2 +\nonumber\\
&& N \left( \frac{\alpha}{2} \,{r}^{D-2} F - \frac{(D-2)}{4} r^{D-3} \eta\,F \, F' \right. \nonumber\\
&& \left. - \frac{(D-3)(D-2)}{4} r^{D-4}\, F^2 \, \eta  \right)=0,
\end{eqnarray*}
and the last equation $E_{\xi}=0$ is
again proportional to the constraint (\ref{cond}) used previously  to obtain
 the solution. As before, at the special point (\ref{alphaeta2}), this system of equations is equivalent to our original equations supplemented by the constraint (\ref{cond}) which also appears as a field equation. The most general solution yields to (\ref{sol2}) together with $N(r)=1$.

 We are now in position to compute the variation
\begin{eqnarray}
\delta B_E=-\beta\mbox{Vol}(\Sigma_{D-2})(D-2)r^{D-3}\times\nonumber\\
\left[\delta F\left(1+\frac{3}{4}\eta
F\xi\right)+\frac{\eta}{2}F^2\delta\xi\right]_{r_+}^{\infty},
\end{eqnarray}
which permits to obtain that
$$
I_E=\frac{(D-1)(D-2)-2l^2\Lambda}{2(D-1)}\left[\beta M-4\pi r_+^{D-2}\right],
$$
We derive the mass ${\cal M}=\frac{\partial I_E}{\partial \beta}$ and the entropy ${\cal S}=\beta\frac{\partial I_E}{\partial \beta}-I_E$, that read
\begin{eqnarray}
&&{\cal M}=\left[\frac{(D-1)(D-2)-2l^2\Lambda}{2(D-1)}\right]M\mbox{Vol}(\Sigma_{D-2})\\
&&{\cal S}=\left[\frac{(D-1)(D-2)-2l^2\Lambda}{2(D-1)}\right]4\pi r_+^{D-2}\mbox{Vol}(\Sigma_{D-2}),\nonumber
\end{eqnarray}
and once again, one can easily check that the first law holds. For the Schwarzschild-AdS case, that is for $\Lambda=-\frac{(D-1)(D-2)}{2l^2}$, these formula reduces to those found in \cite{Birmingham:1998nr}. Finally, the Noether conserved quantity
\begin{eqnarray}
C(r)=Nr^{D-3}(D-2)\Big[\left(1+\frac{3}{4}\eta F\xi\right)\left(-2F+rF^{\prime}\right)\nonumber\\
+\frac{\eta}{2}F^2(r\xi^{\prime}+2\xi)\Big],
\end{eqnarray}
which is a consequence of the scaling symmetry of the reduced action (\ref{redaction2})
\begin{eqnarray}
\bar{r}=\sigma r,\quad \bar{N}(\bar{r})=\sigma^{1-D}N(r),\nonumber\\
\bar{F}(\bar{r})=\sigma^{2}F(r),\quad\bar{\xi}(\bar{r})=\sigma^{-2}\xi(r),
\end{eqnarray}
permits to derive the following Smarr formula
\begin{eqnarray}
{\cal M}=\frac{1}{D-1}\,T\,{\cal S}.
\end{eqnarray}
One may mention that the previous scaling symmetry will not be possible in the case of spherical or hyperboloid horizon.

\section{Conclusions}
We have been concerned with a particular truncation of the Horndeski
theory in three dimensions given by the Einstein-Hilbert piece plus
a cosmological constant and a scalar field with its usual kinetic
term and a nonminimal kinetic coupling. For this model, we have
derived the most general solution subjected to the condition (\ref{cond}).
In this case, the metric turns to be the BTZ spacetime and the
radial scalar field is shown to be well-defined at the horizon. We have seen that such solution occurs because the stress tensor associated to
the variation of the matter source behaves on-shell as a cosmological constant term. The
constraint on the radial component of the conserved current (\ref{cond}) together
with the fact of looking for a static scalar field only impose a
restriction on the geometry and not on the scalar field. In other
words, this means that the field equation associated to the
variation of the scalar field is automatically satisfied without
imposing any restriction on the radial dependence of the scalar
field. In order to compute the mass and the entropy of the solution,
we have constructed an Euclidean action whose field equations turn
out to be equivalent to the original Einstein equations and such
that the constraint on the radial component of the conserved current
appears also as a field equation. This last fact has resulted to be
primordial to obtain the mass and the entropy, and we have verified
that the first law was satisfied. This reduced action has also be
useful in order to derive the usual Smarr formula by exploiting a
scaling symmetry. We have extended all these results in arbitrary dimension where the metric solution is
nothing but the Schwarzschild-AdS spacetime with a planar horizon. In this case also, we have been able to construct an Euclidean
action whose field equations are equivalent to the original ones supplemented by the constraint
condition on the geometry (\ref{cond}). In all these examples, the horizon topology is planar but this hypothesis is only essential in order to establish
a scaling symmetry of the reduced action. In fact, in the spherical or hyperboloid cases, one would be able to construct along the same lines the Euclidean action sharing the same
features except enjoying the scaling symmetry. In higher dimensions, the authors of Ref.
\cite{Anabalon:2013oea} have considered the same model and found black hole solutions with spherical and hyperboloid horizon topology. In these cases, they
have computed the thermodynamical
quantities by regularizing the action with the use of a regular
soliton solution. It will be interesting to see wether
the Euclidean approach described here may yield the same results. It is also appealing that up to now all the known solutions of the
equations (\ref{eqs}) are those where the constraint
(\ref{cond}) is imposed. It will be interesting to  see
wether there exist other solutions for which the radial component of the
current conservation is not vanishing. Finally, we have also
obtained a particular time-dependent solution where the scalar field
depends linearly on the time. For such solution, the issue
concerning the thermodynamical analysis is not clear for us. Hence,
a natural work will consist in providing a consistent Hamiltonian
formalism in order to compute the mass, the entropy and also to give
a physical interpretation on the additional constant $q$ that
appears in the solution. This problem will also be relevant in the context of the Lifshitz case where the known solutions
\cite{Bravo-Gaete:2013dca,Kobayashi:2014eva} are necessarily time-dependent.

\begin{acknowledgments}
We thank Julio Oliva for useful discussions. MB is supported by BECA
DOCTORAL CONICYT 21120271. MH is partially supported by grant
1130423 from FONDECYT and from CONICYT, Departamento de Relaciones
Internacionales ``Programa Regional MATHAMSUD 13 MATH-05''.
\end{acknowledgments}


\end{document}